\documentclass[12pt,letterpaper]{amsart}
\pdfoutput=1
\usepackage[latin1]{inputenc}
\usepackage[bookmarks=false,bookmarksopen=false]{hyperref}
\usepackage{cite,color,eucal,graphicx}
\newcommand{\myetal}{et al.\mbox{}}

\newcommand{\mysize}[1]{{\lvert #1 \rvert}}

\newcommand{\myeqdot}{.}
\newcommand{\myv}[1]{\mathbf{#1}}
\newcommand{\mybound}[2]{\Delta^{#1}_{#2}}

\newcommand{\myutil}{\omega}
\newcommand{\myalg}{\mathcal{A}}

\newcommand{\myE}{\mathcal{E}}
\newcommand{\myH}{\mathcal{H}}
\newcommand{\myQ}{\mathcal{Q}}
\newcommand{\myT}{\mathcal{T}}

\newcommand{\myinstA}{{\mathcal{S}}}
\newcommand{\myinstB}{{\mathcal{S}'}}
\newcommand{\myHA}{\myH}
\newcommand{\myHB}{{\myH'}}
\newcommand{\myVA}{V}
\newcommand{\myVB}{V'}
\newcommand{\myIA}{I}
\newcommand{\myIB}{I'}
\newcommand{\myKA}{K}
\newcommand{\myKB}{K'}

\newtheorem{theorem}{Theorem}

\newtheorem{corollary}[theorem]{Corollary}

\theoremstyle{definition}

\theoremstyle{remark}

\hypersetup{
    colorlinks=false,
    pdfborder={0 0 0},
    pdftitle={Approximating max-min linear programs with local algorithms},
    pdfauthor={Patrik Floréen, Petteri Kaski, Topi Musto, Jukka Suomela},
}

\begin{document}

\title[]{Approximating max-min linear programs \\
    with local algorithms}

\author[]{Patrik Floréen}
\author[]{Petteri Kaski}
\author[]{Topi Musto}
\author[]{Jukka Suomela}

\address{%
    Helsinki Institute for Information Technology HIIT \\
    University of Helsinki \\
    Department of Computer Science \\
    P.O. Box 68 \\
    FI-00014 University of Helsinki \\
    Finland
}
\email{$\{$firstname.lastname$\}$@cs.helsinki.fi}
\thanks{This research was supported in part by the Academy of Finland, Grants 116547 and 117499, and by Helsinki Graduate School in Computer Science and Engineering (Hecse).}

\begin{abstract}
    A local algorithm is a distributed algorithm where each node must operate solely based on the information that was available at system startup within a constant-size neighbourhood of the node. We study the applicability of local algorithms to max-min LPs where the objective is to maximise $\min_k \sum_v c_{kv} x_v$ subject to $\sum_v a_{iv} x_v \le 1$ for each $i$ and $x_v \ge 0$ for each $v$. Here $c_{kv} \ge 0$, $a_{iv} \ge 0$, and the support sets $V_i = \{ v : a_{iv} > 0 \}$, $V_k = \{ v : c_{kv}>0 \}$, $I_v = \{ i : a_{iv} > 0 \}$ and $K_v = \{ k : c_{kv} > 0 \}$ have bounded size. In the distributed setting, each agent $v$ is responsible for choosing the value of $x_v$, and the communication network is a hypergraph $\mathcal{H}$ where the sets $V_k$ and $V_i$ constitute the hyperedges. We present inapproximability results for a wide range of structural assumptions; for example, even if $|V_i|$ and $|V_k|$ are bounded by some constants larger than 2, there is no local approximation scheme. To contrast the negative results, we present a local approximation algorithm which achieves good approximation ratios if we can bound the relative growth of the vertex neighbourhoods in $\mathcal{H}$.
\end{abstract}

\maketitle

\section{Introduction}

We study the limits of what can and what cannot be achieved by \emph{local algorithms}~\cite{naor95what}. We focus on the \emph{approximability} of a certain class of linear optimisation problems, which generalises beyond widely studied packing LPs; the emphasis is on deterministic algorithms and worst-case analysis.

\subsection{Local algorithms}

A local algorithm is a distributed algorithm where each node must operate solely based on the information that was available at system startup within a constant-size neighbourhood of the node. We focus on problems where the size of the input per node is bounded by a constant; in such problems, local algorithms provide an extreme form of scalability: the communication, space and time complexity of a local algorithm is constant per node, and a local algorithm scales to an arbitrarily large or even infinite network.

The study of local algorithms has several uses beyond providing highly scalable distributed algorithms. The existence of a local algorithm shows that the function can be computed by \emph{bounded-fan-in, constant-depth Boolean circuits}; we can say that the function is in the class NC$^0$. A local algorithm is also an efficient centralised algorithm: the time complexity of the centralised algorithm is \emph{linear} in the number of nodes; furthermore, due to spatial locality in memory accesses, we may be able to achieve a low I/O complexity in the \emph{external memory}~\cite{vitter01external} model of computation. In certain problems, a local approximation algorithm can be used to construct a \emph{sublinear time} algorithm which approximates the size of the optimal solution, assuming that we tolerate an additive error and some probability of failure~\cite{parnas07approximating}. A local algorithm can be turned into an efficient \emph{self-stabilising} algorithm~\cite{dolev00self-stabilization}; the time to stabilise is constant \cite{awerbuch91distributed}. Finally, the existence and nonexistence of local algorithms gives us insight into the \emph{algorithmic value of information} in distributed decision-making~\cite{papadimitriou91value}.

\subsection{Max-min packing problem}

In this section, we define the optimisation problem that we study in this work. Let $V$, $I$ and $K$ be index sets with $I \cap K = \emptyset$; we say that each $v \in V$ is an \emph{agent}, each $k \in K$ is a \emph{beneficiary party}, and each $i \in I$ is a \emph{resource} (\emph{constraint}). We assume that one unit of activity by $v$ benefits the party $k$ by $c_{kv}\geq 0$ units and consumes $a_{iv}\geq 0$ units of the resource $i$; the objective is to set the activities to provide a fair share of benefit for each party. In notation, assuming that the activity of agent $v$ is $x_v$ units, the objective is to
\begin{equation}
    \begin{aligned}
        &\text{maximise } & \myutil = \min_{k \in K} \sum_{v \in V} c_{kv} x_v & \\
        &\text{subject to } & \sum_{v \in V} a_{iv} x_v &\le 1 & \text{for each } &i \in I, \\
        && x_v &\ge 0 & \text{for each } &v \in V \myeqdot
    \end{aligned}\label{eq:max-min}
\end{equation}

Throughout this work we assume that the \emph{support sets} defined 
for all $i\in I$, $k\in K$, and $v\in V$ by $V_i=\{v\in V:a_{iv}>0\}$, $V_k=\{v\in V:c_{kv}>0\}$, $I_v = \{ i \in I : a_{iv}>0\}$, and $K_v = \{ k \in K : c_{kv}>0 \}$ have bounded size. That is, we consider only instances of~\eqref{eq:max-min} such that $\mysize{I_v} \le \mybound IV$, $\mysize{K_v} \le \mybound KV$, $\mysize{V_i} \le \mybound VI$ and $\mysize{V_k} \le \mybound VK$ for some constants $\mybound IV$, $\mybound KV$, $\mybound VI$ and $\mybound VK$. To avoid uninteresting degenerate cases, we furthermore assume that $I_v$, $V_i$ and $V_k$ are nonempty.

\subsection{LP formulation}

If the sets $V$, $I$ and $K$ are finite, the problem can be represented using matrix notation. Let $A$ be the nonnegative $\mysize{I} \times \mysize{V}$ matrix where the entry at row $i$, column $v$ is $a_{iv}$; define $C$ analogously. We write $a_i$ for the row $i$ of $A$ and $c_k$ for the row $k$ of $C$. Let $x$ be a column vector of length $\mysize{V}$. The goal is to maximise $\myutil = \min_{k \in K} c_k x$ subject to $A x \le \myv{1}$ and $x \ge \myv{0}$.

In the special case $\mysize{K}=1$, this is the widely studied fractional packing problem: maximise $c x$ subject to $A x \le \myv{1}$ and $x \ge \myv{0}$. This simple linear program (LP) has nonnegative coefficients in $c$ and $A$. We refer to a problem of this form as a \emph{packing LP}; the dual is a \emph{covering LP}. Naturally the case of any finite $K$ can also be written as a linear program, but the constraint matrix is no longer nonnegative: maximise $\myutil$ subject to $A x \le \myv{1}$, $\myutil \myv{1} - Cx \le \myv{0}$ and $x \ge \myv{0}$.

\subsection{Distributed setting}\label{ssec:distributed-setting}

We construct the hypergraph $\myH = (V, \myE)$ where the hyperedges are $\myE = {\{ V_i : i \in I \}} \, \cup \, {\{ V_k : k \in K \}}$. This is the \emph{communication graph} in our distributed optimisation problem. The variable $x_v$ is controlled by the agent $v \in V$, and two agents $u, v \in V$ can communicate directly with each other if they are adjacent in $\myH$. We write $d_\myH(u,v)$ for the shortest-path distance between $u$ and $v$ in $\myH$. The agents are cooperating, not selfish; the difficulty arises from the fact that the agents have to make decisions based on incomplete information.

Initially, each agent $v \in V$ knows only the following local information: the identity of its neighbours in the graph $\myH$; the sets $I_v$ and $K_v$; the values $a_{iv}$ for each $i \in I_v$; and the values $c_{kv}$ for each $k \in K_v$. That is, $v$ knows with whom it is competing on which resources, and with whom it is working together to benefit which parties.

When we compare the present work with previous work, we often mention the special case $\mysize{K} = 1$, as this corresponds to the widely studied packing LP. However, in this case the size of $V_k$ is not bounded by a constant $\mybound VK$: we have $V_k = V$ for the sole $k \in K$. Therefore we introduce a restricted variant of the distributed setting, which we call \emph{collaboration-oblivious}. In this variant, the hyperedges are $\myE = {\{ V_i : i \in I \}}$. Whenever we study related work on the packing LP, we focus on the collaboration-oblivious setting.

\subsection{Local setting}\label{ssec:local-setting}

We are interested in solving the problem \eqref{eq:max-min} by using a local algorithm. Let $r=1,2,\ldots$ be the \emph{local horizon} of the algorithm; this is a constant which does not depend on the particular problem instance at hand. Let $B_\myH(v, r)=\{u\in V:d_\myH(u,v)\leq r\}$ be the set of nodes which have distance at most $r$ to the node $v$ in $\myH$. The agent $v$ must choose the value $x_v$ based on the information that is initially available in the agents $B_\myH(v, r)$.

We focus on the case where the size of the input is constant per node. The elements $a_{iv}$ and $c_{kv}$ are represented at some finite precision. Furthermore, we assume that the nodes have constant-size locally unique identifiers; i.e., any node can be identified uniquely within the local horizon.

\subsection{Approximation}

A local algorithm has the \emph{approximation ratio} $\alpha$ for some $\alpha > 1$ if the decisions $x_v$ are a feasible solution and the value $\myutil$ is within a factor $\alpha$ of the global optimum. A family of local algorithms is a \emph{local approximation scheme} if we can achieve any $\alpha > 1$ by choosing a large enough local horizon~$r$.

\subsection{Contributions}

In Section~\ref{sec:inapprox} we show that while a simple algorithm achieves the approximation ratio $\mybound VI$ for \eqref{eq:max-min}, no local algorithm can achieve an approximation ratio less than $\mybound VI/2 + 1/2 - 1/(2 \mybound VK - 2)$ in the general case. In Section~\ref{sec:approx} we present a local approximation algorithm which can achieve an improved approximation ratio if we can bound the relative growth of the vertex neighbourhoods in $\myH$.

\section{Applications}

Consider a two-tier sensor network: battery-powered sensor devices generate some data; the data is transmitted to a battery-powered relay node, which forwards the data to a sink node. The sensor network is used to monitor the physical areas $K$. Let $S$ be the set of sensors, and let $T$ be the set of relays; choose $I = S \cup T$.

For each sensors device $s \in S$, there may be multiple relays $t \in T$ which are within the reach of the radio of $s$; we say that there is a wireless link $(s,t)$ from $s$ to $t$. The set $V$ consists of all such wireless links, and the variable $x_{(s,t)}$ indicates how much data is transmitted from $s$ via $t$ to the sink. Transmitting one unit of data on the link $v = (s,t) \in V$ and forwarding it to the sink consumes the fraction $a_{sv}$ of the energy resources of the sensor $s$ and also the fraction $a_{tv}$ of the energy resources of the relay~$t$.

Let $c_{kv} = 1$ for each link $v = (s,t)$ if the sensor $s$ is able to monitor the physical area $k \in K$. Now \eqref{eq:max-min} captures the following optimisation problem: choose the data flows in the sensor network so that we maximise the minimum amount of data that is received from any physical area. Equivalently, we can interpret the objective as follows: choose data flows such that the lifetime of the network (time until the first sensor or relay runs out of the battery) is maximised, assuming that we receive data at the same average rate from each physical area.

Similar constructions have applications beyond the field of sensor networks: consider, for example, the case where each $k \in K$ is a major customer of an Internet service provider (ISP), each $s \in S$ is a bounded-capacity last-mile link between the customer and the ISP, and each $t \in T$ is a bounded-capacity access router in the ISP's network.

\section{Related work}

Papadimitriou and Yannakakis~\cite{papadimitriou93linear} present the \emph{safe algorithm} for the packing LP. The agent $v$ chooses
\begin{equation}\label{eq:safe}
    x_v = \min_{i \in I_v} \frac{1}{a_{iv} \mysize{V_i}} \myeqdot
\end{equation}
This is a local $\mybound VI$-approximation algorithm with horizon $r = 1$.

Kuhn \myetal{}~\cite{kuhn06price} present a distributed approximation scheme for the packing LP and covering LP. The algorithm provides a local approximation scheme for some families of packing and covering LPs. For example, let $a_{iv}\in\{0,1\}$ for all $i, v$. Then for each $\mybound VI$, $\mybound IV$ and $\alpha > 1$,  there is a local algorithm with some constant horizon $r$ which achieves an $\alpha$-approximation. Our work shows that such local approximation schemes do not exist for~\eqref{eq:max-min}.

Another distributed approximation scheme by Kuhn \myetal{}~\cite{kuhn06price} forms several decompositions of $\myH$ into subgraphs, solves the optimisation problem optimally for each subgraph, and combines the solutions. However, the algorithm is not a local approximation algorithm in the strict sense that we use here: to obtain any constant approximation ratio, the local horizon must extend (logarithmically) as the number of variables increases. Also Bartal \myetal{}~\cite{bartal97global} present a distributed but not local approximation scheme for the packing LP.

Kuhn and Wattenhofer~\cite{kuhn05constant-time} present a family of local, constant-factor approximation algorithms of the covering LP that is obtained as an LP relaxation of the minimum dominating set problem. Kuhn \myetal{}~\cite{kuhn05locality} present a local, constant-factor approximation of the packing and covering LPs in unit-disk graphs.

There are few examples of local algorithms which approximate linear problems beyond packing and covering LPs. Kuhn \myetal{}~\cite{kuhn06fault-tolerant} study an LP relaxation of the $k$-fold dominating set problem and obtain a local constant-factor approximation for bounded-degree graphs.

For combinatorial problems, there are both negative~\cite{kuhn04what,linial92locality} and positive~\cite{floreen07local,kuhn06fault-tolerant,kuhn05constant-time,naor95what,urrutia07local} results on the applicability of local algorithms.

\section{Inapproximability}\label{sec:inapprox}

Even though the safe algorithm~\cite{papadimitriou93linear} was presented for the special case of $\mysize{K} = 1$, $c = \myv 1$, and finite $I$ and $V$, we note that the safe solution $x$ defined by~\eqref{eq:safe} and the optimal solution $x^{*}$ also satisfy
\[
    \min_{k \in K} \sum_{v \in V_k} c_{kv} x^{*}_v
    \le \min_{k \in K} \sum_{v \in V_k} c_{kv} \mybound VI x_v
    = \mybound VI \min_{k \in K} \sum_{v \in V_k} c_{kv} x_v \myeqdot
\]
Therefore we obtain a local approximation algorithm with the approximation ratio $\mybound VI$ for~\eqref{eq:max-min}.

One could hope that widening the local horizon beyond $r = 1$ would significantly improve the quality of approximation. In general, this is not the case: no matter what constant local horizon $r$ we use, we cannot improve the approximation ratio beyond $\mybound VI / 2$. In this section, we prove the following theorem.

\begin{theorem}\label{thm:inapprox}
    Let $\mybound VI \ge 2$ and $\mybound VK \ge 2$ be given. There is no local approximation algorithm for \eqref{eq:max-min} with the approximation ratio less than $\mybound VI/2 + 1/2 - 1/(2 \mybound VK - 2)$. This holds even if we make the following restrictions: ${a_{iv}\in\{0,1\}}$, ${\mybound IV = 1}$ and ${\mybound KV = 1}$.
\end{theorem}

We emphasise that the local algorithm could even choose any local horizon $r$ depending on the bounds $\mybound VI$, $\mybound VK$, $\mybound IV$ and $\mybound KV$. Nevertheless, an arbitrarily low approximation ratio cannot be achieved if ${\mybound VI \ge 3}$ or ${\mybound VK \ge 3}$. In the case $\mybound VI = \mybound VK = 2$ the existence of a local approximation scheme remains an open question.

Analogous proof techniques, using constructions based on regular bipartite high-girth graphs, have been applied in previous work to prove the local inapproximability of packing and covering LPs~\cite{kuhn06price} and combinatorial problems~\cite{kuhn04what}.

\subsection{Proof outline}

Choose any local approximation algorithm $\myalg$ for the problem \eqref{eq:max-min}. Let $r \ge 1$ be the local horizon of $\myalg$ and let $\alpha$ be the approximation ratio of~$\myalg$. We derive a lower bound for $\alpha$ by constructing two instances of \eqref{eq:max-min}, $\myinstA$ and $\myinstB$, such that certain sets of nodes in the two instances have identical radius-$r$ neighbourhoods in both instances. Consequently, the deterministic local algorithm $\myalg$ must make the same choices for these nodes in both instances. The nodes with identical views are selected based on the solution of $\myinstA$ computed by $\myalg$, which enables us to obtain a lower bound on $\alpha$ by showing that this solution is necessarily suboptimal as a solution of $\myinstB$. 

\begin{figure}
    \rotatebox{0.001}{\scalebox{0.8}{\input{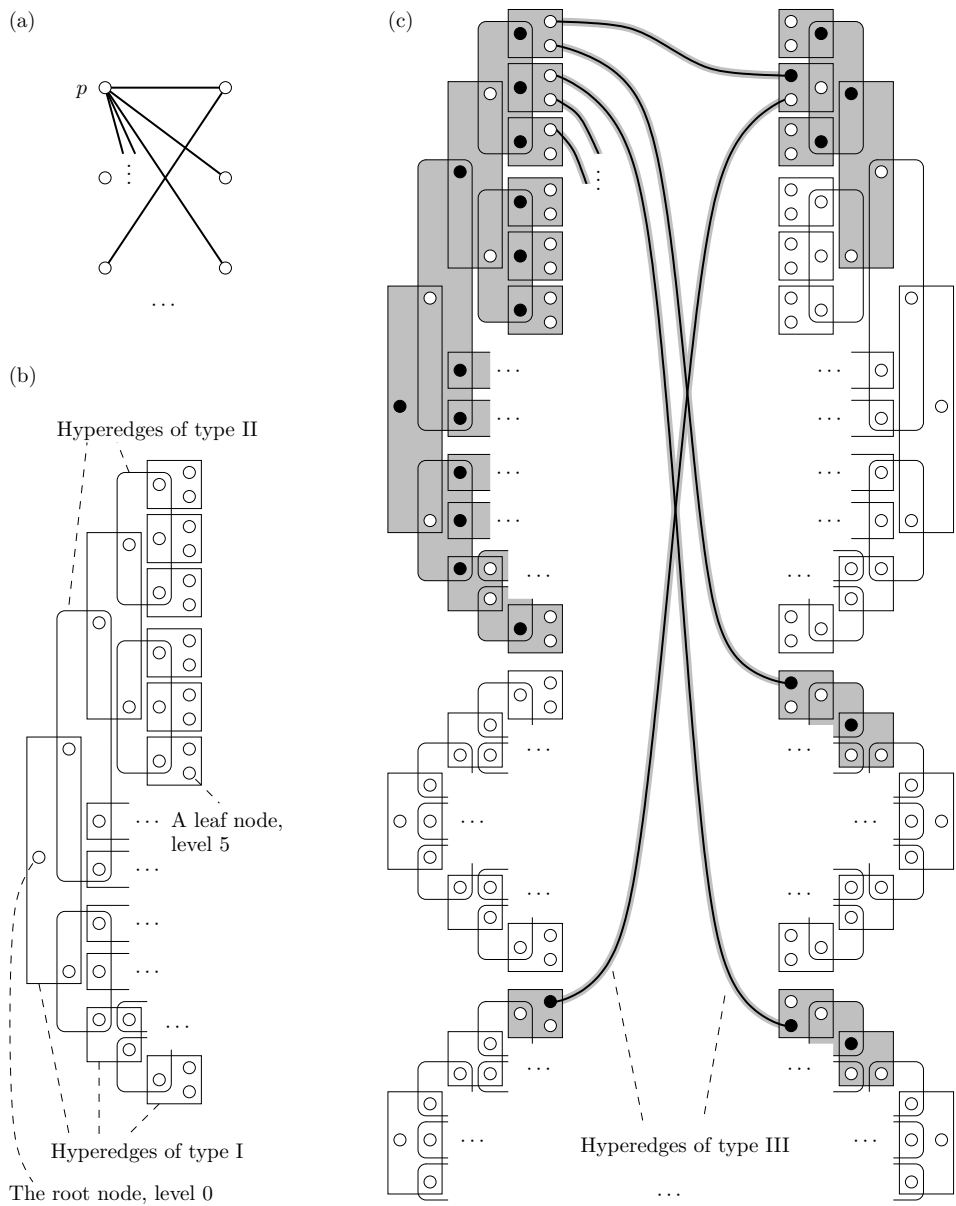}}}
    \caption{The construction of $\myinstA$, in the case $d = 2$, $D = 3$, $r = 2$, $R = 3$.
    (a)~A small part of the bipartite, 72-regular, high-girth graph $\myQ$.
    (b)~A complete $(2,3)$-ary hypertree of height~$5$, with 72 leaves.
    (c)~The underlying hypergraph of $\myinstA$.
    Grey highlighting indicates the underlying hypergraph of $\myinstB$; black circles are the variables of $\myinstB$ which we set to $1$ in Section~\ref{ssec:inapprox-feasible-T}.}\label{fig:inapprox}
\end{figure}

\subsection{Construction of $\myinstA$}
We now proceed with the detailed construction of the instance $\myinstA$. The constructions used in the proof are illustrated in Figure~\ref{fig:inapprox}.

Let $\mybound VI \ge 2$ and $\mybound VK \ge 2$; without loss of generality we can assume that at least one of the inequalities is strict because setting $\mybound VI=\mybound VK=2$ in the theorem statement yields the trivial bound $\alpha\geq 1$. Let $d = \mybound VI - 1$ and $D = \mybound VK - 1$. Observe that $dD>1$. Let $R>r$; the precise value of $R$ is chosen later and will depend on $d$, $D$ and $\alpha$ only. 

Let $\myQ$ be a $d^R D^{R-1}$-regular bipartite graph with no cycles consisting of less than $4r+2$ edges. (A random regular bipartite graph with sufficiently many nodes has this property with positive probability~\cite{mckay04short}.) The graph $\myQ$ provides the template for constructing the hypergraph underlying the instance $\myinstA$. 

Before describing the construction, we first introduce some terminology. A \emph{complete $(d,D)$-ary hypertree} of \emph{height} $h$ is defined inductively as follows. For $h=0$, the hypertree consists of exactly one node and no edges; the \emph{level} of the node is $0$. For $h>0$, start with a complete $(d,D)$-ary hypertree of height $h-1$. For each node $v$ at level $h-1$, introduce a new hyperedge and new nodes as follows. If $h-1$ is even, the new hyperedge consists of the node $v$ and $d$ new nodes. If $h-1$ is odd, the new hyperedge consists of the node $v$ and $D$ new nodes. For future reference, call these hyperedges of types I and II, respectively. The new nodes have level $h$ in the constructed hypertree. The constructed hypertree is a complete $(d,D)$-ary hypertree of height $h$. The \emph{root} of the hypertree is the node at level 0, the \emph{leaves} are the nodes at level $h$. Each level $\ell$ has either $(dD)^{\ell/2}$ or $(dD)^{(\ell-1)/2}d$ nodes depending on whether $\ell$ is even or odd, respectively. See Figure~\ref{fig:inapprox} for an illustration. 

We now construct the hypergraph underlying $\myinstA$. Denote by $Q$ the vertex set of $\myQ$. Form a hypergraph $\myH$ by taking $\mysize{Q}$ node-disjoint copies of a complete $(d,D)$-ary hypertree of height $2R-1$. For $q\in Q$, denote the copy corresponding to $q$ by $\myT_q$. Denote the node set of $\myT_q$ by $T_q$. For $\ell=0,1,\ldots,2R-1$, denote the set of nodes at level $\ell$ in $\myT_q$ by $T_q(\ell)$. Denote the set of leaf nodes in $\myT_q$ by $L_q=T_q(2R-1)$.

Observe that the number of leaf nodes in each $\myT_q$ is equal to the degree of every vertex in $\myQ$. For each vertex $q\in Q$ and each leaf node $v\in L_q$, associate with $v$ a unique edge of $\myQ$ incident with the vertex $q$. Each edge of $\myQ$ is now associated with exactly two leaf nodes; by construction, these leaf nodes always occur in different hypertrees $\myT_q$. For a leaf $v\in\cup_q L_q$, let $f(v)$ be the other leaf associated with the same edge of $\myQ$. Observe that $f(f(v))=v$ holds for all $v\in\cup_q L_q$; in particular, $f$ is a permutation of $\cup_q L_q$. To complete the construction of $\myH$, add the hyperedge $\{v,f(v)\}$ to $\myH$ for each $v\in\cup_q L_q$. Call these hyperedges type III hyperedges.

Let us now define the instance $\myinstA$ of \eqref{eq:max-min} based on the hypergraph $\myH$. Let the set of agents $\myVA$ be the node set of $\myH$. For each hyperedge $e$ of type I, there is a resource $i \in \myIA$; let $a_{iv}=1$ if $v \in e$, otherwise $a_{iv}=0$. For each hyperedge $e$ of type II, there is a beneficiary party $k \in \myKA$; let $c_{kv}=1/D$ if $v \in e$, otherwise $c_{kv}=0$. For each hyperedge $e$ of type III, there is a beneficiary party $k \in \myKA$; let $c_{kv}=1$ if $v \in e$, otherwise $c_{kv}=0$. The locally unique identifiers of the agents can be chosen in an arbitrary manner. (This proof applies also if the identifiers are globally unique;  for example, we can equally well consider the standard definition where the identifiers are a permutation of $1,2,\ldots,\mysize{\myVA}$.) This completes the construction of $\myinstA$. Observe that $\myinstA$ has $\myH$ as its underlying hypergraph.

\subsection{Construction of $\myinstB$}

Next we construct another instance of \eqref{eq:max-min}, called $\myinstB$, by restricting to a part of $\myinstA$. To select the part, we apply the algorithm $\myalg$ to the instance $\myinstA$. We do not care what is the optimal solution of $\myinstA$; all that matters at this point is the fact that each agent $v \in \myVA$ must choose some value $x_v \ge 0$. In particular, we pay attention to the values $x_v$ at the leaf nodes $v\in\cup_q L_q$.

For all $q\in Q$, let
\begin{equation}
\label{eq:delta-q}
    \delta(q)=\sum_{v\in L_q}(x_v-x_{f(v)}) \myeqdot
\end{equation}
For all $P\subseteq Q$, let $\delta(P)=\sum_{q\in P}\delta(q)$. Because $f$ is a permutation of $\cup_q L_q$ with $f(f(v))=v$ for all $v\in\cup_q L_q$, we have $\delta(Q) = 0$. Thus, there exists a $p\in Q$ with $\delta(p)\geq 0$.

The instance $\myinstB$ is now constructed based on $p$. The set of agents in $\myinstB$ is
\[
    \myVB = T_p\ \cup\ \bigcup_{u \in L_p} B_\myH(u, 2r),
\]
the set of resources is $\myIB = \{ i \in \myIA : V_i \subseteq \myVB \}$, and the set of beneficiary parties is $\myKB = \{ k \in \myKA : V_k \subseteq \myVB \}$. The coefficients $a_{iv}$ and $c_{kv}$ for $i \in \myIB$, $k \in \myKB$, and $v\in \myVB$ are the same as in the instance $\myinstA$. The locally unique identifiers of the agents $v \in \myVB$ are the same as in the instance $\myinstA$. (If we prefer globally unique identifiers which are a permutation of $1,2,\ldots,\mysize{\myVB}$, we can add redundant variables to $\myVB$.)

\subsection{The structure of $\myinstB$}

Next we show that the structure of $\myinstB$ is tree-like, that is, there are no cycles in the hypergraph $\myHB$ defined by the instance $\myinstB$; by construction, $\myHB$ is a subgraph of $\myHA$.

For each $q\in Q$, the subgraph induced by $T_q$ in $\myHA$ is a hypertree. Furthermore, the subsets $T_q$ form a partition of $\myVA$. Therefore any cycle in $\myHA$ and, therefore, any cycle in $\myHB$ must involve hyperedges which cross between the subsets $T_q$ and, finally, return back to the same subset.

The only hyperedges which connect nodes in $T_q$ and $T_{w}$ for distinct $q,w\in Q$ are the hyperedges of type III. There is at most one such hyperedge for any fixed $q\neq w$; this hyperedge corresponds to the edge $\{q,w\}$ in the graph $\myQ$. Therefore a cycle in $\myHB$ implies a cycle in $B_\myQ(p, 2r)$; this implies a cycle of length at most $4r+1$ in $\myQ$; by construction, no such cycle exists.

\subsection{A feasible solution of $\myinstB$}\label{ssec:inapprox-feasible-T}

Next we show that there is a feasible solution $\hat{x}$ of $\myinstB$ with $\myutil = 1$. Let $u$ be the root node in $\myT_p$. By construction, $u\in T_p\subseteq \myVB$. For each $v\in \myVB$, let $\hat{x}_v = 1$ if $d_\myHB(u,v)$ is even; otherwise, let $\hat{x}_v = 0$. See Figure~\ref{fig:inapprox} for an illustration.

Because $\myinstB$ is tree-like, there is a unique path connecting $u$ to $v$ in $\myHB$ for each $v\in \myVB$. In particular, this path is a shortest path and has length $d_\myHB(u,v)$. It follows that each hyperedge in $\myHB$ has a unique node (that is, the node having the minimum distance to $u$) of its distance parity to $u$. Observe that hyperedges of resources and beneficiary parties alternate in paths from $u$. By the structure of $\myinstA$ and $\myinstB$, it follows that the hyperedges of resources (type I) have a unique node with even distance to $u$. Therefore, $\sum_{v\in \myVB} a_{iv} \hat{x}_v = 1$ for each $i\in \myIB$; the solution is feasible. Analogously, the hyperedges of beneficiary parties (types II and III) have a unique node with odd distance to $u$. Therefore, $\sum_{v \in \myVB} c_{kv} \hat{x}_v = 1$ for each $k\in \myKB$, implying $\myutil = 1$.

\subsection{The solution achieved by $\myalg$ in $\myinstB$}

Now we apply $\myalg$ to $\myinstB$. The local radius-$r$ view of the nodes $v \in T_p$ is identical in both $\myinstA$ and $\myinstB$. In particular, the deterministic local algorithm $\myalg$ must make the same choices $x_v$ for $v\in T_p$ in both instances.

As there is a feasible solution with $\myutil = 1$, the approximation algorithm $\myalg$ must choose a solution $x$ with $\sum_{v \in \myVB} c_{kv} x_v \allowbreak\ge 1/\alpha$ for all~$k\in \myKB$.

We proceed in levels $\ell=0,1,\ldots,2R-1$ of $\myT_p$. 
We study the \emph{total} value assigned to the variables at level $\ell$, 
defined by
\[
    S(\ell) = \sum_{v \in T_p(\ell)} x_v \myeqdot
\]
Recall that $|T_p(\ell)|=(dD)^{l/2}$ for $\ell$ even, and
$|T_p(\ell)|=(dD)^{(l-1)/2}d$ for $\ell$ odd.

Let us start with level $\ell=2R-1$, that is, the leaf nodes in $\myT_p$. For each $v \in L_p$, there is a $k \in \myKB$ such that $\myVB_k = \{ v, f(v) \}$ and $c_{kv} = c_{k f(v)} = 1$. Therefore,
by \eqref{eq:delta-q} and the fact that $\delta(p) \ge 0$,
\begin{equation}
\label{eq:s-2r-1}
    S(2R-1) = \sum_{v \in L_p} x_v
    = \frac12 \delta(p) + \frac12 \sum_{v \in L_p} \left(  x_v + x_{f(v)} \right)
    \ge \frac{d^R D^{R-1}}{2\alpha} \myeqdot
\end{equation}

Next, we study the remaining odd levels $\ell=2j-1$ for $j=1,2,\ldots,\allowbreak R-1$. Consider the set $F_p(2j-1)=T_p(2j-1)\cup T_p(2j)$. Observe that the hyperedges of type II occurring in $F_p(2j-1)$ form a partition of $F_p(2j-1)$. Each of the $d^jD^{j-1}$ hyperedges in the partition has exactly one node in $T_p(2j-1)$ and exactly $D$ nodes in $T_p(2j)$. The coefficients $c_{kv}$ of each beneficiary party $k\in \myKB$ associated with these hyperedges are $1/D$ for all $v \in \myVB_k$. Thus, by the approximation ratio, we obtain the bound
\begin{equation}
\label{eq:s-2j-1-2j}
    S(2j-1)+S(2j)
    = \sum_{k\in \myKB:\, \myVB_k\subseteq F_p(2j-1)} D \sum_{v \in \myVB_k} c_{kv} x_v
    \ge d^jD^j / \alpha \myeqdot
\end{equation}

Let us finally study the even levels $\ell=2j$ for $j=0,1,2,\ldots,R-1$. Observe that the hyperedges of type I occurring in $F_p(2j)=T_p(2j)\cup T_p(2j+1)$ partition $F_p(2j)$. Each of the $d^jD^j$ hyperedges in the partition has exactly one node in $T_p(2j)$ and exactly $d$ nodes in $T_p(2j+1)$. The coefficients $a_{iv}$ of the resources $i\in \myIB$ associated with these hyperedges are $1$ for all $v \in \myVB_i$. Thus, by the feasibility of $x$, we obtain the bound
\begin{equation}
\label{eq:s-2j-2j+1}
    S(2j) + S(2j+1)
    = \sum_{i\in \myIB:\, \myVB_i\subseteq F_p(2j)} \  \sum_{v \in \myVB_i} a_{iv} x_v
    \le d^jD^j \myeqdot
\end{equation}

Put together, we have, for $j=1,2,\ldots,R-1$,
\begin{align}
    \label{eq:s-1-0-1}
    S(1) &\le S(0) + S(1) \le 1, \\
    \label{eq:s-2j-1-2j+1}
    S(2j-1) &\overset{\text{\eqref{eq:s-2j-1-2j}}}{\ge} d^jD^j / \alpha - S(2j)
    \overset{\text{\eqref{eq:s-2j-2j+1}}}{\ge} S(2j+1) - \left(1-\frac{1}{\alpha} \right) d^jD^j
\end{align}
which, together with the assumption $d D > 1$, implies
\begin{align*}
    1 \overset{\text{\eqref{eq:s-1-0-1}}}{\ge} S(1)
    &\overset{\text{\eqref{eq:s-2j-1-2j+1}}}{\ge} S(2R-1) - \left(1-\frac{1}{\alpha} \right) \sum_{j=1}^{R-1} d^j D^j \\
    &\overset{\text{\eqref{eq:s-2r-1}}}{\ge} \frac{d^R D^{R-1}}{2\alpha} - \left(1-\frac{1}{\alpha} \right) \frac{d^R D^R - d D}{d D - 1} \myeqdot
\end{align*}
Therefore 
$
    \alpha \ge d/2 + 1 - 1/(2D) + {(d + 2 - 2 d D - 1/D)} /\allowbreak {(2 d^R D^R - 2)}
$.
Should we have $\alpha < d/2 + 1 - 1/(2D)$, we would obtain a contradiction by choosing a large enough $R$. This concludes the proof of Theorem~\ref{thm:inapprox}.

The same proof with $D=1$ gives the following corollary which shows inapproximability even if both $a_{iv}\in\{0,1\}$ and $c_{kv}\in\{0,1\}$.
\begin{corollary}\label{cor:inapprox}
    Let $\mybound VI > 2$ be given. There is no local approximation algorithm for \eqref{eq:max-min} with the approximation ratio less than $\mybound VI/2$. This holds even if we make the following restrictions: ${a_{iv}\in\{0,1\}}$, ${c_{kv}\in\{0,1\}}$, ${\mybound VK = 2}$, ${\mybound IV = 1}$ and ${\mybound KV = 1}$.
\end{corollary}

\section{Approximability}\label{sec:approx}

We have seen that the approximation ratio provided by the safe algorithm is within factor $2$ of the best possible in general graphs; there is no local approximation scheme if $\mybound VI > 2$ or $\mybound VK > 2$.

However, the graph in our construction is very particular: it is tree-like, and the number of nodes in a radius-$r$ neighbourhood grows exponentially as the radius $r$ increases. Such properties are hardly realistic in practical applications such as sensor networks; if nodes are embedded in a low-dimensional physical space, the length of each communication link is bounded by the limited range of the radio, and the distribution of the nodes and the network topology are not particularly pathological, we expect that the number of nodes grows only polynomially as the radius $r$ increases. We shall see that better approximation ratios may be achieved in such cases.

Formally, we define the relative growth of neighbourhoods by
\[
    \gamma(r) = \max_{v \in V} \frac{\mysize{B_\myH(v, r+1)}}{\mysize{B_\myH(v, r)}} \myeqdot
\]
We prove the following theorem.
\begin{theorem}\label{thm:approx}
    For any $R$, there is a local approximation algorithm for \eqref{eq:max-min} with the approximation ratio $\gamma(R-1) \, \gamma(R)$ and local horizon~$\Theta(R)$.
\end{theorem}

To illustrate this result, consider the case where $\myH$ is a $d$-dimensional grid. In such a graph, we have $\mysize{B_\myH(v, r)} = \Theta(r^d)$ and $\mysize{B_\myH(v, r+1)} = \mysize{B_\myH(v, r)} + \Theta(r^{d-1})$. Therefore $\gamma(r) = 1 + \Theta(1/r)$ and our algorithm is a local approximation scheme in this family of graphs.

We emphasise that the algorithm does not need to know any bound for $\gamma(r)$. We can use the same algorithm in any graph. The algorithm achieves a good approximation ratio if such bounds happen to exist, and it still produces a feasible solution if such bounds do not exist. Furthermore, due to the local nature of the algorithm, if the graph fails to meet such bounds in a particular area, this only affects the optimality of the beneficiary parties that are close to this area.

\subsection{Algorithm}

The algorithm is based on the idea of averaging local solutions of local LPs; similar ideas have been used in earlier work to derive distributed and local approximation algorithms for LPs~\cite{kuhn07distributed,kuhn05locality,kuhn06price}.

\begin{figure}
    \scalebox{0.8}{\input{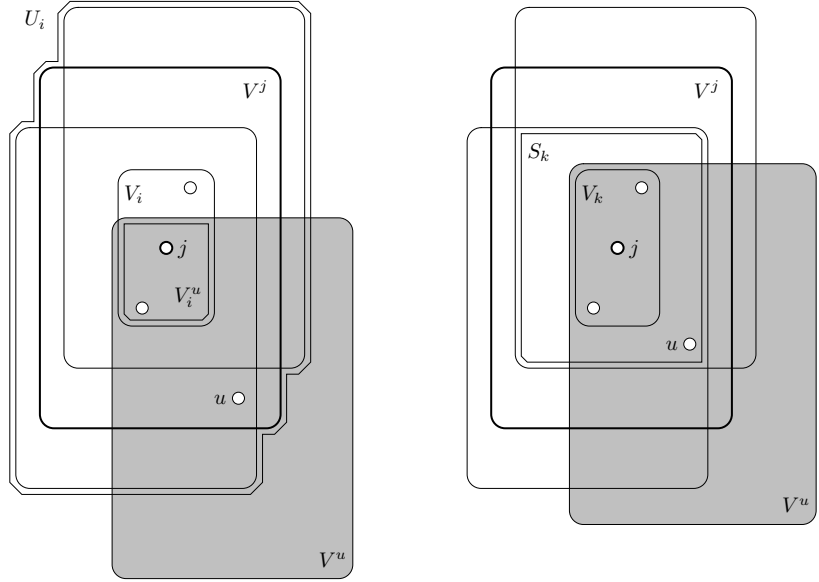}}
    \caption{Definitions used in the algorithm.}\label{fig:approx}
\end{figure}

Fix a radius $R=1,2,\ldots$; the local horizon of the algorithm will be $\Theta(R)$. For each agent $u \in V$, define
\begin{align*}
    V^u &= B_\myH(u, R), &
    K^u &= \{ k \in K : V_k \subseteq V^u \}, \\
    V_i^u &= V_i \cap V^u, &
    I^u &= \{ i \in I : V_i^u \ne \emptyset \} \myeqdot
\end{align*}
For each $k \in K$ and $i \in I$, define
\begin{align*}
    S_k &= \!\bigcap_{j \in V_k}\! V^j, &
    m_k &= \mysize{S_k}, &
    M_k &= \max {\{ \mysize{V^j} : j \in V_k \}}, \\
    U_i &= \!\bigcup_{j \in V_i}\! V^j, &
    N_i &= \mysize{U_i}, &
    n_i &= \min {\{ \mysize{V^j} : j \in V_i \}} \myeqdot
\end{align*}
See Figure~\ref{fig:approx} for an illustration. For each $u \in V$, let $x^u$ be an optimal solution of the following problem:
\begin{equation}
    \begin{aligned}
        &\text{maximise } & \myutil^u = \min_{k \in K^u} \sum_{v \in V_k} c_{kv} x^u_v & \\
        &\text{subject to } & \sum_{v \in V_i^u} a_{iv} x^u_v &\le 1 & \text{for each } &i \in I^u, \\
        && x^u_v &\ge 0 & \text{for each } &v \in V^u \myeqdot
    \end{aligned}\label{eq:max-min-local}
\end{equation}
The solution $x^u$ can be computed by the agent $u$; or it can be computed separately by each agent $j \in V^u$ which needs $x^u$, by using the same deterministic algorithm.

The agent $j \in V$ makes the following choice, which depends only on its radius $2R+1$ neighbourhood:
\begin{equation}\label{eq:alg-choice}
    \beta_j = \min_{i\in I_j} \frac{n_i}{N_i}, \qquad
    \tilde{x}_j = \frac{\beta_j}{\mysize{V^j}} \sum_{u \in V^j} x^u_j \myeqdot
\end{equation}

\subsection{Constraints}

Consider a resource $i \in I$. We note that
\begin{equation}\label{eq:const1}
\begin{aligned}
    j \in V_i \text{ and } u \in V^j &\iff u \in U_i  \text{ and } j \in V_i \text{ and }  u \in V^j \\
        &\iff u \in U_i \text{ and } j \in V_i \text{ and }  j \in V^u \\
        &\iff u \in U_i \text{ and } j \in V_i^u
\end{aligned}
\end{equation}
and
\begin{equation}\label{eq:const2}
\begin{aligned}
    u \in U_i &\implies \exists j \in V_i : u \in V^j
        \iff \exists j \in V_i : j \in V^u \\
        &\iff V_i^u \ne \emptyset
        \iff i \in I^u
        \overset{\text{\eqref{eq:max-min-local}}}{\implies} \sum_{v \in V^u_i} a_{iv} x^u_v \le 1 \myeqdot
\end{aligned}
\end{equation}
By definition, $\beta_j \le n_i / N_i$ for all $i \in I_j$, that is, for all $j \in V_i$. Combining these observations, we obtain
\begin{align*}
    \sum_{j \in V_i} a_{ij} \tilde{x}_j
    &\overset{\text{\eqref{eq:alg-choice}}}{=} \sum_{j \in V_i} a_{ij} \frac{\beta_j}{\mysize{V^j}} \sum_{u \in V^j} x^u_j
    \le \frac{1}{n_i} \frac{n_i}{N_i} \sum_{j \in V_i} \, \sum_{u \in V^j} a_{ij} x^u_j \\
    &\overset{\text{\eqref{eq:const1}}}{=} \frac{1}{N_i} \sum_{u \in U_i} \, \sum_{j \in V_i^u} a_{ij} x^u_j
    \overset{\text{\eqref{eq:const2}}}{\le} \frac{1}{N_i} \sum_{u \in U_i} 1 = 1,
\end{align*}
Therefore $\tilde{x}$ is a feasible solution of \eqref{eq:max-min}.

\subsection{Benefit}

Let $x^{*}$ be an optimal solution of \eqref{eq:max-min}, with $\myutil = \myutil^{*}$. Then $x^{*}$ is a feasible solution of \eqref{eq:max-min-local}, with $\myutil^u \ge \myutil^{*}$. Therefore the optimal solution $x^u$ of \eqref{eq:max-min-local} satisfies
\begin{equation}\label{eq:ben0}
    \sum_{v \in V_k} c_{kv} x^u_v \ge \myutil^{*}
\end{equation}
for all $k \in K^u$. Let $\beta = \min_{j \in V} \beta_j = \min_{i\in I} n_i / N_i$.

Consider a beneficiary party $k \in K$. We note that
\begin{equation}\label{eq:ben1}
    u \in S_k \text{ and } j \in V_k \implies j \in V_k \text{ and } u \in V^j
\end{equation}
and
\begin{equation}\label{eq:ben2}
\begin{aligned}
    u \in S_k &\implies u \in V^j \text{ for all } j \in V_k
        \iff j \in V^u \text{ for all } j \in V_k \\
        &\iff V_k \subseteq V^u
        \iff k \in K^u
        \overset{\text{\eqref{eq:ben0}}}{\implies} \sum_{v \in V_k} c_{kv} x^u_v \ge \myutil^{*} \myeqdot
\end{aligned}
\end{equation}
Combining these observations, we obtain
\begin{align*}
    \sum_{j \in V_k} c_{kj} \tilde{x}_j
    &\overset{\text{\eqref{eq:alg-choice}}}{=} \sum_{j \in V_k} c_{kj} \frac{\beta_j}{\mysize{V^j}} \sum_{u \in V^j} x^u_j
    \ge \frac{\beta}{M_k} \sum_{j \in V_k} \, \sum_{u \in V^j} c_{kj} x^u_j \\
    &\overset{\text{\eqref{eq:ben1}}}{\ge} \frac{\beta}{M_k} \sum_{u \in S_k} \, \sum_{j \in V_k} c_{kj} x^u_j
    \overset{\text{\eqref{eq:ben2}}}{\ge} \frac{\beta}{M_k} \sum_{u \in S_k} \myutil^{*} = \beta \, \frac{m_k}{M_k} \, \myutil^{*} \myeqdot
\end{align*}

In summary, the solution $\tilde{x}$ approximates \eqref{eq:max-min} within the approximation ratio $\max_{k \in K} M_k/m_k \cdot \max_{i \in I} N_i/n_i$. To complete the proof of Theorem~\ref{thm:approx}, observe that $\max_{k \in K} M_k / m_k \le \gamma(R-1)$ and $\max_{i \in I} N_i / n_i \le \gamma(R)$. 

\bibliographystyle{abbrv}
\bibliography{articles}

\end{document}